\documentclass{svjour3}                     
\smartqed  
\usepackage{graphicx,amssymb,amsmath}
\def\be{\begin{eqnarray} &&}
\def\nonu{\nonumber \\ &&}
\def\ee{\end{eqnarray}}
\def\psla{\rlap \slash}
\def\bew{\begin{widetext}}
\def\ew{\end{widetext}}

\usepackage{color}

\renewcommand{\bar}[1]{\overline{#1}}

\begin{document}

\title{
Exploring the Pion phenomenology within a fully 
covariant
constituent quark model
\thanks{The present research activity is partially supported by the Italian
MIUR grant PRIN 2008.}
}

\titlerunning{Exploring the Pion phenomenology  }

\author{Giovanni Salm\`e, Emanuele Pace and  Giovanni Romanelli 
}

\authorrunning{Giovanni Salm\`e et al}

\institute{ 
           Giovanni Salm\`e \at
Istituto Nazionale di Fisica Nucleare,
Sezione di Roma, P.le A. Moro 2,
I-00185, Italy \\
Tel.: +39-6-49914872\\
              Fax: +39-6-4454749\\
	      \and
             E. Pace \at 
	     Phys. Dept., ''Tor Vergata'' University and INFN Sezione
	      di Tor Vergata, Via della Ricerca Scientifica 1, 00133 Rome, Italy
               \\
	       \and G. Romanelli \at Phys. Dept., ''Tor Vergata''
	        University, Via della Ricerca Scientifica 1, 00133 Rome, Italy
	      }

\date{Received: date / Accepted: date}

\maketitle

\begin{abstract}
An overview of the  vector and tensor Generalized Parton Distributions 
for a charged pion is presented.  Such
 observables,   belonging to the set of quantities fundamental for a detailed investigation
 of  
 the hadronic inner dynamics,  have been evaluated within a fully  covariant Constituent 
 Quark Model, based on
a proper Ansatz of the pion Bethe-Salpeter Amplitude  and the Mandelstam formula
for matrix elements of  operators acting on relativistic composite systems.
Given the very encouraging results already obtained for  the vector distribution 
of a
charged pion, 
 the model has been extended to the tensor one, and some
preliminary calculations will be illustrated.
\end{abstract}

\section{Introduction}
\label{intro}
The 
non perturbative regime of Quantum Chromodynamics (QCD), namely
the  present theory of the strong interaction, is 
  still a paramount challenge. In particular,  achieving a complete
    3D description of  hadrons is the topic of 
  many and  coherent efforts, both on experimental and theoretical sides.
The perturbative regime, with its fundamental feature, the asymptotic
freedom, has been  experimentally  investigated in great detail, and this 
has allowed to access the short-distance behavior of the hadronic wave
functions. But, in order to shed light on  confinement, the other 
peculiar feature of QCD  (deeply related to  its non linearity), one has to move
towards the external part of the hadron wave function.  

From the theoretical point of view, the  Light-front (LF) framework, with
variables defined by: $a^\pm=a^0\pm
a^3$ and  ${\bf a}_\perp\equiv \{a_x,a_y\}$,  is very
suitable  for describing  relativistic, interacting systems, like
hadrons. Among the several motivations for adopting the LF framework, beyond the
well-known  fact that the dynamics of the light-cone is
naturally described in terms of  LF variables,  one has to mention 
that the vacuum is
almost trivial within a LF field theory \cite{Brodpr}. Therefore, e.g.
 for the pion, one
can construct the following meaningful Fock expansion (with some caveat for the
zero-mode contributions \cite{Brodpr})
$$| \pi \rangle =  ~|q\bar{q} \rangle ~~ + 
|q \bar{q} ~ q \bar{q}\rangle ~ + ~
|q \bar{q} ~g\rangle .....$$
where the valence component $|q\bar{q} \rangle$ and the non valence ones are
fully considered.
Within our approach, based on a covariant description of the pion state
(see \cite{Frede09,Frede11} and references quoted therein), one can
explore also the contributions beyond the valence term.

In  recent years, it has been  
 recognized that 
a wealth of information on the partonic structure of hadrons is encoded in 
 the Generalized Parton
Distributions (GPD) (see, e.g., Ref. \cite{Diehlpr}),  as well as  in 
the Transverse-momentum 
Distributions (TMD) (see,
e.g., Ref.  \cite{Baronpr}).
Those observables, that can be experimentally accessed through Deep Virtual
Compton Scattering and Semi-inclusive Deep Inelastic Scattering, respectively,
 are expressed by quantities invariant for LF-boosts, and
allow one to parametrize matrix elements (between hadronic states) of relevant
quark-quark or gluon-gluon correlators. Aim of our covariant, phenomenological
analysis is the evaluation of the leading-twist pion GPD's.
\section{Generalized Parton
Distributions}
The  generalized parton
distributions are off-diagonal (respect to the hadron four-momenta, i.e.
 $p_f\ne p_i$) matrix elements of  quark-quark (or gluon-gluon) 
correlator projected onto the  Dirac basis (see,
e.g., Ref. \cite{Meis08} for a thorough investigation of the pion case), 
and naturally summarize a lot of information, scattered in many
observables, like e.g.  electromagnetic (em) form factors or  
parton-distribution functions. 
In particular,  the pion, i.e. the simplest hadron  to be considered, 
 has two leading-twist quark GPD's: i) the vector, or no spin-flip, GPD,
 $H^I_\pi(x,\xi,t)$,
 and ii) the tensor, or
spin-flip, GPD, $ E^I_{\pi,T}(x,\xi,t)$ ($I=IS,IV$ labels 
 isoscalar
and isovector GPD's).
In order to avoid
 Wilson-line contributions, one
  can choose the light-cone gauge \cite{Diehlpr}
 and gets 
\be
2~\left(\begin{array}{c} 
 H^{IS}_{\pi}(x, \xi, t) \\ ~\\ H^{IV}_{\pi}(x, \xi, t) \end{array}\right)
 =
  \int {dk^- d{\bf k}_\perp\over 2}\nonu
  \times
 \int \frac{dz^- dz^+ d{\bf z}_\perp}{2(2 \pi)^4} 
 e^{i[(xP^+z^-+k^-z^+)/2 -{\bf z}_\perp \cdot {\bf k}_\perp]}~\langle p'| 
\bar{\psi}_q (-\frac12z) \gamma^+ \left(\begin{array}{c} 1 \\ ~\\\tau_3
  \end{array}\right) \psi_q(\frac12z) 
|p \rangle
=
\nonu= 
 \int \frac{dz^- }{4 \pi} e^{i(xP^+z^-)/2}\langle p'| 
\bar{\psi}_q (-\frac12z) \gamma^+ \left(\begin{array}{c} 1 \\ ~\\\tau_3
  \end{array}\right) \psi_q(\frac12z) 
|p \rangle\big|_{\tilde z=0}
\label{vector}\ee
and
\be
\frac{P^+ \Delta^j- P^j\Delta^+}{P^+ m_\pi}\left(\begin{array}{c} 
 E^{IS}_{\pi T}(x, \xi, t) \\ ~\\E^{IV}_{\pi T}(x, \xi, t) \end{array}\right)
 =
 \nonu= 
 \int \frac{dz^- }{4 \pi} e^{i(xP^+z^-)/2}\langle p'| 
\bar{\psi}_q (-\frac12z) ~i\sigma^{+j} ~\left(\begin{array}{c} 1 \\ ~\\\tau_3
  \end{array}\right) \psi_q(\frac12z) 
|p \rangle\big|_{\tilde z=0}
\label{tensor}\ee
where $\tilde z \equiv \{z^+ = z^0 + z^3 , {\bf z}_\perp\}$,
  $\psi_q(z)$ is the  quark-field  isodoublet and 
  \be
  x={k^+\over P^+}~~, \quad \xi= ~-~{\Delta^+\over 2 P^+}~~, \quad 
  t=-\Delta^2 ~~, \quad  \Delta=p'-p~~, \quad  P= {p'+p\over 2}
  \ee
  with $\{k^+- \Delta^+/2,{\bf k}_\perp-{\bf \Delta}_\perp/2\}$  the initial
   LF momentum of the active quark. Notice that the factor of two multiplying
   the vector GPD is chosen in order to  obtain the 
   em form
   factor from $\int dx~ H(x,\xi,t)$  (see below).
   
   As anticipated above, there exist basic relations between the GPD's and 
   other
  relevant observables. If we consider a charged pion, 
   the vector GPD is related both to the parton distributions,
  $q(x)$,  and to 
  the em form  as
  follows
  \be
  H^u_\pi (x, 0,
 0)=\theta(x)u(x)-\theta(-x) \bar u(-x)\nonu 
 F_\pi(t)= \int_{-1}^1 dx ~ H^{IV}_\pi (x, \xi, t)= \int_{-1}^1 dx ~
  H^{u}_\pi (x, \xi, t)
 \ee
 where $H^u_\pi=H^{IS}_\pi + H^{IV}_\pi$.  
  The relation between the non-spin flip GPD and the em form
 factor of a charged pion
 can be generalized, if one considers  Mellin moments of both vector and tensor
 GPD's, obtaining the so-called Generalized Form Factors (GFF). In particular,
 for the $u$-quark vector and tensor GPD's, 
 the corresponding Mellin moments are given by
\be
 \int_{-1}^1 dx \, x^n H^u_\pi (x, \xi, t) = 
\sum _{i=0}^{\ell} (2 \xi) ^{i} A^u_{n+1,2i} (t)~~, 
\nonu
\int_{-1}^1 dx \, x^n E^u_{\pi,T} (x, \xi, t) = 
\sum _{i=0}^{\ell} (2 \xi) ^{i} B^u_{n+1,2i} (t)  
\label{gff}\ee 
where $\ell=\{(n+1)/2\}$ ($\{a\}$ is the integer part of $a$), 
 $A^u_{n+1,2i}(t)$ is a  vector GFF and   $B^u_{n+1,2i}(t)$ a tensor one.
  One has isoscalar (isovector) GFF's if $n$ is  odd (even).
   It is worth noting that 
 \be \int_{-1}^1 dx \,  H^u_\pi (x, \xi, t)= A^u_{1,0} (t)=F_\pi(t)
 \label{ff}\ee
 and
 \be\int_{-1}^1 dx \,  E^u_{\pi,T} (x, \xi, t)= B^u_{1,0} (t)\ee
 with  $B^u_{1,0} (0)\ne 0$, the tensor charge for $n=1$, or the
   tensor anomalous magnetic moment (as it is called in Ref. \cite{Hagelpr}).

In order to attach a physical interpretation to the GFF, one can generalize 
to a
relativistic framework what is familiar for the em form factors (scalars and
then 
boost-invariant
observables) in a non
relativistic approach. In this case, one interprets the 3D Fourier transforms 
of
the em form factors of a hadron as the intrinsic (Galilean-invariant)  
em distributions in the
coordinate space (e.g., for the pion, 
one has the charge distribution, while, for the nucleon, one has both charge and
magnetic distributions). In the relativistic case, one has to consider  Fourier 
transforms of GPD's, that are quantities
   depending upon variables 
  invariant under LF boosts. Moreover, only the transverse part of $\Delta^\mu$
can be trivially conjugated to coordinate-space variables, while for  $x$ 
and $\xi$ (proportional to
$\Delta^+$) this is not possible. Therefore one can introduce bidimensional
 Fourier transforms with
respect to ${\bf \Delta}_\perp$, for gaining some insight in the spatial
distributions of the quarks, still keeping  the description invariant for  
proper boosts (i.e. LF boosts).  In particular, from Eq. (\ref{gff}) one can see
that, putting $\xi=0$, only $A^u_{n+1,0} (\Delta^2)$ 
(cf also Eq. (\ref{gff}))  and 
$B^u_{n+1,0} (\Delta^2)$ survive. There exist an infinite set of frame where 
$\xi=0$,  due to the LF-invariance of $\xi$, that are indicated as 
the  Drell-Yan frames. In these frames, where 
  $\Delta^+=0$ and  
$\Delta^2=-\Delta^2_\perp$,  one can introduce 
the Fourier transforms of the previously  defined GFF's as follows  
\be
\tilde A_{n}(b_\perp)= \int {d {\bf \Delta}_\perp\over (2 \pi)^2} ~e^{i 
{\bf\Delta}_\perp \cdot {\bf b}_\perp}
  A_{n,0}(\Delta^2)~, ~ 
\tilde B_{n}(b_\perp)= \int {d {\bf \Delta}_\perp\over (2 \pi)^2}
~ e^{i {\bf  \Delta}_\perp \cdot {\bf b}_\perp} 
B_{n,0}(\Delta^2) 
\ee
where $b_\perp=|{\bf b}_\perp |$. 
They have a direct physical interpretation 
as quark densities 
in the impact-parameter space (IPS) \cite{Bur,Bur08}. 
In particular,  $\tilde A_{n}(b_\perp)$ represents the probability density  
of finding an unpolarized quark 
in the pion at a certain distance $b_\perp$ from the transverse
center of momentum. 
If one takes into account the quark polarization, then 
  the probability density of finding in the Drell-Yan frame
 a charged parton of fixed transverse polarization, 
 ${\bf s}_\perp$, and given  ${\bf b}_\perp$, is
\be
\rho_n ({\bf b}_\perp,{\bf s}_\perp) = 
\frac12 \left [ \tilde A_n(b_\perp) + \frac {s^i\epsilon ^{ij} b^j}{b_\perp} ~\Gamma_n(b_\perp)
\right ] 
\label{rhodef}\ee
where
\be
\Gamma_n(b_\perp) = -\frac{1}{2 m_{\pi}}\frac{\partial ~\tilde B_n(b_\perp)}{\partial ~b_\perp}
\label{gamdef}
\ee
 The IPS density $\rho_n$ can be obtained by calculating the  proper 
matrix elements of the projector for  a transversely-polarized quark 
($s^\mu\equiv \{0,0,{\bf s}_\perp\}$), namely
\be { 1\over 2}\gamma^+ \left [1 +{\bf s}_\perp \cdot {\boldsymbol{ \gamma}}_\perp\gamma_5\right]=
{ 1\over 2}\left [\gamma^+ -i\sum_j  s^j  ~\sigma^{+ j}\gamma_5\right]=
{ 1\over 2}\left [\gamma^+ -\sum_{ij}  s^j  ~\sigma^{+ i}\epsilon^{ji}\right]~~.
\ee

It should be pointed out that the distribution density of a quark with a given
helicity (longitudinal polarization) is expressed by   the first term in Eq.
(\ref{rhodef}),
since the pion is a pseudoscalar meson and the quark  helicity
operator (for a massless fermion: $\gamma_5/2$) has a vanishing expectation 
value, due to parity invariance.

Equation (\ref{rhodef}) clearly indicates how to access the quark distribution
in the impact-parameter space through the study of the GPD's. Moreover, as a
closing remark, one could   
exploit $E_{\pi T}$ to extract  more elusive information on the quasi-particle nature
of the constituent quarks, like their possible anomalous magnetic moments,
once the vector current that governs the quark-photon coupling,  is suitably 
improved (see below and \cite{Brom07} for a discussion in the lattice framework).
  \section{Transverse momentum distributions}
   The transverse momentum distributions are  diagonal (in the pion
   four-momentum) matrix elements of  the  
   quark-quark (or gluon-gluon) 
correlator  with the proper Wilson-line contributions (see,
e.g., Ref. \cite{Meis08}) and  suitable Dirac structures. The TMD's depend upon $x$
and the  quark transverse momentum, ${\bf k}_\perp$  (notice that $\Delta^\mu =0$ leads to 
 $\xi=t=0$),  and since   the integration over
 ${\bf k}_\perp$ is not performed, the
Wilson-line effects must be carefully analyzed.  At the leading-twist one
has two TMD's, for the pion: the T-even $f_{1}(x, |{\bf k}_\perp|^2)$,
that yields the probability distribution to find an unpolarized quark with LF momentum
$\{x,{\bf k}_\perp\}$ in the pion, and the  T-odd  
$h^\perp_1(x, |{\bf k}_\perp|^2,\eta)$, 
related to a  transversely-polarized quark and called {\em Boer-Mulders}
distribution \cite{Boer}. The two TMD's allow one to parametrize the distribution
of a quark with given  LF momentum and transverse polarization, i.e.
(see, e.g., Ref. \cite{Brom08})
\be
\rho (x,{\bf k}_\perp,{\bf s}_\perp) = 
\frac12 \left [f_{1}(x, |{\bf k}_\perp|^2)  + 
{s^i\epsilon ^{ij} k_\perp^j \over m_\pi}
 ~h^\perp_1(x, |{\bf k}_\perp|^2,\eta)
\right ] 
\label{rhodefk}
\ee
where the dependence upon the variable $\eta$ in $h^\perp_1$ can be traced back
to  
the    Wilson-line effects.
Moreover, it is worth  reminding that  ${\bf b}_\perp$ is not conjugated to 
${\bf k}_\perp$, but to ${\bf \Delta}_\perp$.

 By choosing the light-cone gauge and the advanced boundary condition 
for the gauge field, the effect of the Wilson lines (final state interaction
effects) can be  translated into complex phases affecting the initial state 
(see, e.g., Ref. \cite{Beli03}).
At the lowest order,  the unpolarized TMD 
$f^{I}_{1}(x, |{\bf k}_\perp|^2)$,  acquaints   the  following form
\be
2\left(\begin{array}{c} 
 f^{IS}_{1}(x, |{\bf k}_\perp|^2) \\ ~\\  f^{IV}_{1}(x, |{\bf k}_\perp|^2) \end{array}\right)
 =
 \nonu= 
 \int {dz^-  d{\bf z}_\perp \over 2 (2\pi)^3}e^{i [xP^+z^-/2 -
 {\bf k}_\perp\cdot
 {\bf z}_\perp] }\langle p| 
\bar{\psi}_q (-\frac12z) \gamma^+ \left(\begin{array}{c} 1 \\ ~\\\tau_3
  \end{array}\right) \psi_q(\frac12z) 
|p \rangle\big|_{ z^+=0}~~~,
\ee
After integrating over ${\bf k}_\perp$, one gets  
  the standard unpolarized parton
distribution $q(x)$, viz
\be
q(x)= \int d{\bf k}_\perp~ f^{q}_{1}(x, |{\bf k}_\perp|^2)= ~
H^{q}_{1}(x, 0,0)~~~.
\ee
The T-odd TMD, $h^\perp_1(x, |{\bf k}_\perp|^2,\eta)$ 
  vanishes at the lowest order in perturbation
theory,  since it becomes proportional to the matrix elements
\be
\langle p| 
\bar{\psi}_q (-\frac12z) i~\sigma^{+j}~ \left(\begin{array}{c} 1 \\ ~\\\tau_3
  \end{array}\right) \psi_q(\frac12z) 
|p \rangle\big|_{ z^+=0}~~~,
\ee
 that are equal to zero, due to  the time-reversal invariance.
In order to get a non vanishing Boer-Mulders distribution,
one has to  evaluate at least a first-order correction,  involving Wilson 
lines 
  (see, e.g., Refs. \cite{Lu04} and  \cite{Meis08}).

\section{The Covariant Constituent Quark Model}

Presently, 
extensive theoretical and experimental research programs 
are being pursued to gain information on both GPD's and
TMD's of hadrons (particularly of the nucleon). Within such efforts, 
considering
that the lattice calculations (in Euclidean space) cannot cover the whole kinematics that experiments
can investigate, it seems worth considering the development of
phenomenological models, that are able to include    general features,
 like the covariance, and to properly describe both the experimental data and the
 lattice ones. In this way one could have an effective tool for the analysis of 
 the results of the forthcoming experiments (e.g., the ones planned and approved at TJLAB), 
 devoted to set-up the 3D tomography of hadrons.
 
The main ingredients in our covariant constituent quark model (CCQM) are represented
 by i) a  model of
 the 4D quark-hadron  vertex, namely the Bethe-Salpeter
amplitude, and ii) the extension to the  GPD's and TMD's of 
  the 4D Mandelstam expression \cite{Mandel}, originally introduced for
  calculating   matrix elements of the em current
  operator for a relativistic  interacting system.
 It is important to emphasize that our investigation, naturally goes beyond 
 a purely valence description of the pion \cite{Frede09,Frede11}.

\subsection{The Mandelstam Formula for the electromagnetic current}

The Mandelstam formula allows one to express  the
matrix elements of the em current of a composite bound system, within a field
theoretic approach \cite{Mandel}.  It has been applied for evaluating
the form factors of both  pion \cite{Melo06}
and nucleon \cite{Melo09}. Furthermore, it has been  exploited for calculating the
vector GPD of the pion  \cite{Frede09,Frede11} and for a preliminary evaluation of the
tensor GPD \cite{Pace12}.

 For instance, in the case of the em form factor of  the pion,
 in the spacelike region, the Mandelstam formula  within CCQM reads
  (see, e.g., Ref. \cite{Melo06})
\be
j^{\mu} ~=~ ~-\imath 2 e \frac{m^2}{f^2_\pi} N_c
 \int
\frac{d^4k}{(2\pi)^4}
\overline \Lambda_{\pi}(k+q,P'_\pi) 
\Lambda_\pi(k,P_\pi) \nonu \times~  
Tr[S(k-P_\pi)~ \gamma^5
S(k+q)~\Gamma^\mu(k,q)~S(k)~  \gamma^5  ]  
\label{current1}\ee
where $f_\pi=92.4$ MeV  is the pion decay constant, $N_c=3$ is the  number of
colors, $m=220$ MeV the CQ mass,
  $\displaystyle
S(p)=\frac{1}{\psla p-m+\imath \epsilon} $ 
 the fermionic propagator,   and 
$\Gamma^\mu(k,q)$  the quark-photon vertex, that  simplifies to
$\gamma^\mu$ in the spacelike region. In presence of a CQ, one
could  add to the bare  vector current  a term proportional to an anomalous magnetic moment, namely a term
like $$ i{\kappa_q \over 2 m_q}\sigma^{\mu\nu} \Delta_\nu~~~,$$
that should not be confused with the quark contribution to the nucleon anomalous
magnetic moment.
In  the present approach we have disregarded such a possibility (cf
\cite{Brom07}
for an improved vector current  within a lattice framework).
In Eq.
(\ref{current1}), it has been assumed a pion  Bethe-Salpeter amplitude (BSA) 
with the following form
\be
\Psi (k-P,p) =  -{m \over f_\pi}~S\left(k-\Delta/ 2\right)~  \gamma^5 ~
\Lambda(k-P,p)~
S\left(k-P\right) \label{bsa}
\ee
where only the most important Dirac structure $ \gamma_5$ has been retained (see 
Ref. \cite{maris} for a detailed discussion of terms beyond $\gamma^5$), and 
$\Lambda(k-P,p)$ is the momentum-dependent part.

The generalization to the case of GPD's, can be found in Ref. \cite{Frede09,Frede11}
for the vector GPD, and in \cite{Pace12,PRS12} for the tensor one.
In particular, both an accurate description of the em form factor 
of the charged pion for 
$t\leq 0$ \cite{Frede09,Frede11}
and  a nice behavior of the parton distribution evolved up to the experimental
scale (see \cite{PRS12}) can be achieved by adopting the following  analytic 
covariant Ansatz for the momentum-dependent 
part of the BSA  
 \be
 \Lambda(k-P,p)=  
{C \over \left[\left(k-\Delta/2\right)^2-m^2_{R} + \imath \epsilon\right]
\left[\left(P-k \right)^2-m^2_{R}+ \imath \epsilon\right]} 
\label{vertexp}
\ee
where 
the parameter $m_R=1200$ MeV is adjusted  to fit   $f_{\pi}$, while 
 the constants  $C$ is fixed through the form factor normalization,
 $F_{\pi}(t=0)=1$,
 that acts as
 a normalization condition for the BSA. 

 For the sake of concreteness, let us give 
 both vector and tensor GPD's   for the $u$ quark, viz
\be
 2 ~H^u(x,\xi,t) = -\imath {\cal R} ~
\int
\frac{d^4k}{(2\pi)^4} ~ \delta[P^+x-k^+] ~
 \Lambda(k-P,p^{\prime})\; \;
\Lambda(k-P,p)\nonu \times   
Tr\left \{ S\left({k}-{P}\right)
\gamma^5 ~S\left({k}+\Delta/2
\right) 
\gamma^+~S\left({k}-\Delta/2
\right)\gamma^5\right \}
 \label{trace}
\ee
and 
\be
{P^+ \Delta^j- P^j\Delta^+ \over  ~ P^+m_\pi}
 E^u_{\pi T}(x, \xi, t) = i {\cal R}
\int \frac{d^4 k }{(2 \pi)^4} \delta[P^+x-k^+]~\Lambda(k-P,p')~ \Lambda(k-P,p) \nonu
\times 
 Tr[S({k}-{P})\gamma^5S({k}+\Delta/2)\gamma^+\gamma^j 
 S({k}-\Delta/2)\gamma^5 ]~
\ee
where
${\cal R}=2 N_c m^2/f^2_\pi$ and $j=1,2$.
The $\delta$ function imposes the correct support for the
{\em active} quark, i.e.  
$|\xi| \le x \le 1$ (see \cite{Frede09,Frede11}), corresponding to the so-called DGLAP region 
\cite{dglap}.
The diagrammatic representation  of the vector GPD, with a LF-time ordering, is given in Fig. 1a and 1b. 
\begin{figure*}[thb!]
\hspace{-.0cm}\includegraphics[width=5.5cm]{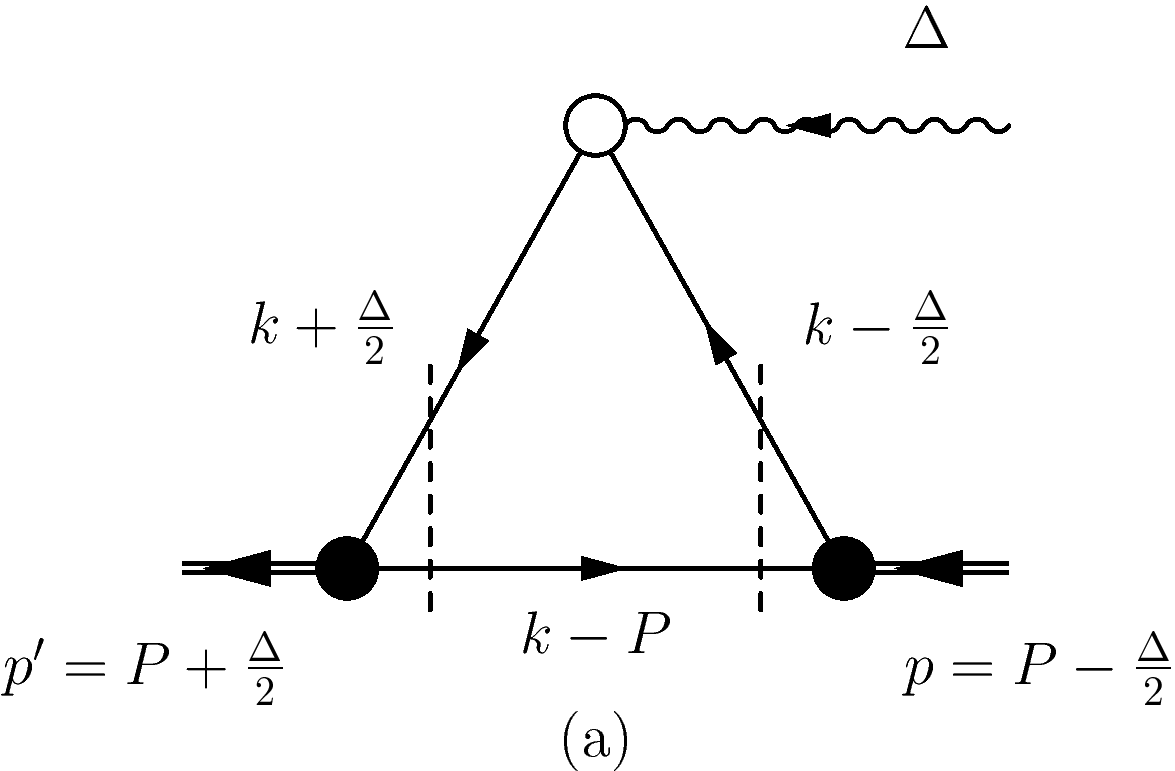}~~~\hspace{0.7cm}
\includegraphics[width=5.5cm]{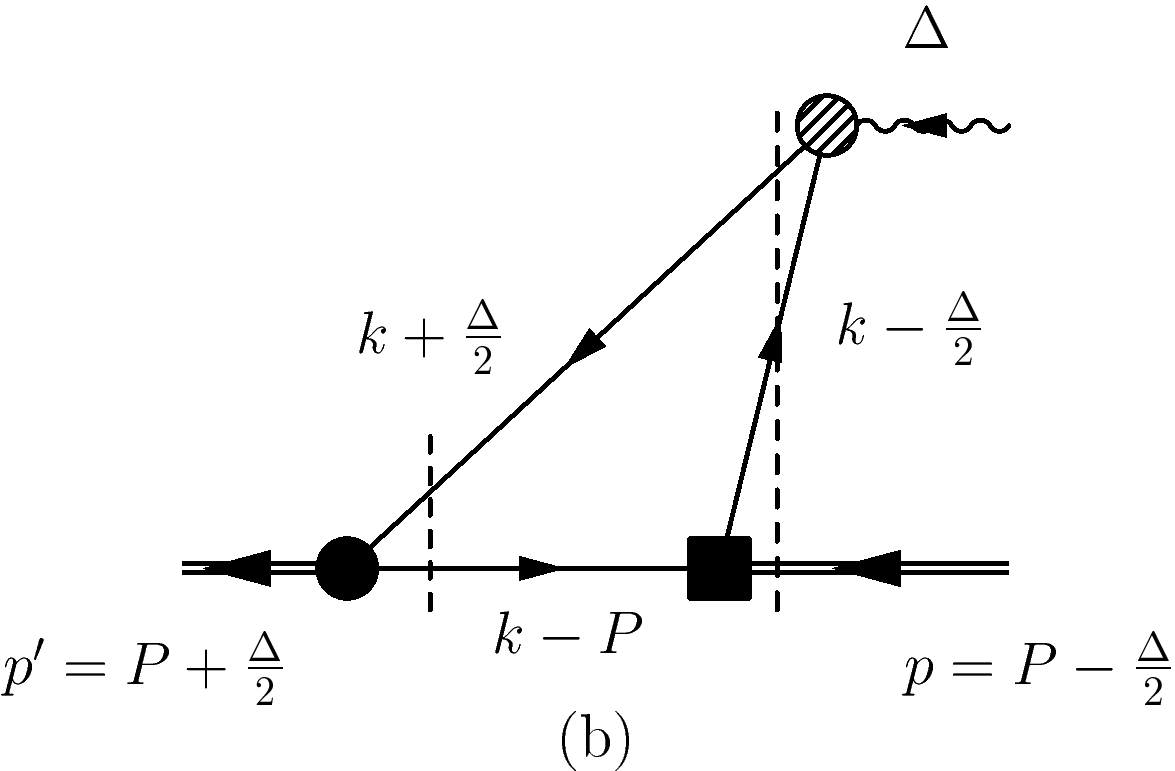}
\caption{LF time-ordered analysis of the pion GPD within the CCQM framework. 
Diagram (a): contribution 
of the active-quark in the valence region with $1\ge x \ge|\xi|$.
Diagram (b):  contribution from the pair-production mechanism
in the nonvalence region, $|\xi|> x >-|\xi|$.
(After Ref. \cite{Frede09,Frede11}).}
\label{figbs} 
\end{figure*} 

In the CCQM investigation of GPD's, a Breit frame with 
$\Delta^+ = -\Delta^- \ge 0$ is chosen in order to   access the whole
range of the variable $\xi$, i.e. $-1\leq \xi \leq 1$. In this way both the
valence and non valence region can be investigated, addressing the interesting
topic of the smooth transition from the DGLAP (valence) regime to the ERBL (non
valence) one \cite{erbl}. In Ref. \cite{Frede09,Frede11}, the vector GPD and the
unpolarized TMD, resulting from
 the
CCQM approach,  are thoroughly 
analyzed. In particular, the CCQM
calculation was able to carefully reproduce the experimental data of the em 
form factor of the charged pion over
 the whole  spacelike range of $t$, presently explored. As to  the  GFF's 
 obtained
from the vector GPD (cf Eq. (\ref{gff})), only lattice data exist and  
solely for $A_{2,0}(t)$ and  
$A_{2,2}(t)$ \cite{Bromth}. Also for those GFF's, the comparison was quite
encouraging. 

In Fig. 2, our preliminary results for the tensor GPD of a charged pion
are shown for some values of the variable $\xi$ and $t$, but for $0\leq x\leq
1$. The GPD for negative values of $x$ can be obtained by exploiting the
 symmetry property
of $E^{IS,IV}_{\pi T}(x,\xi,t)$ (see, e.g. Ref. \cite{Diehlpr}) 
under the 
transformation  $x\to -x$. The symmetry property follows from both  the charge-conjugation ($p\to -p$ and
$p^\prime \to -p^\prime$)
 and  the isospin invariance and leads to
\be
E^{IS}_{\pi T}(x,\xi,t) = -E^{IS}_{\pi T}(-x,\xi,t)~~, \quad \quad
E^{IV}_{\pi T}(x,\xi,t) = E^{IV}_{\pi T}(-x,\xi,t)
\label{sym}\ee
It is important to remind  that for $\xi \to 0$ the valence component is dominant
(DGLAP regime) while for $\xi \to 1$ the non valence term is acting (ERBL
regime). In view of that, it is expected a peak around $x\sim 1$ for
 $\xi \to 1$, as discussed in Ref. \cite{Frede09,Frede11}. Finally, from Eq.
 (\ref{sym}),
 one can understand the vanishing value of 
  $E^{IS}_{\pi T}(0,\xi,t)$. In Ref. \cite{Pace12}, our preliminary
   CCQM results have been compared   with the
  calculations obtained within a LF Hamiltonian Dynamics framework, where only
  the valence component have been retained \cite{Frede11}, obtaining a nice
  agreement in the relevant kinematical regions.
\begin{figure}

\vspace{-3cm}
\begin{center}
\includegraphics[width=6.0cm]{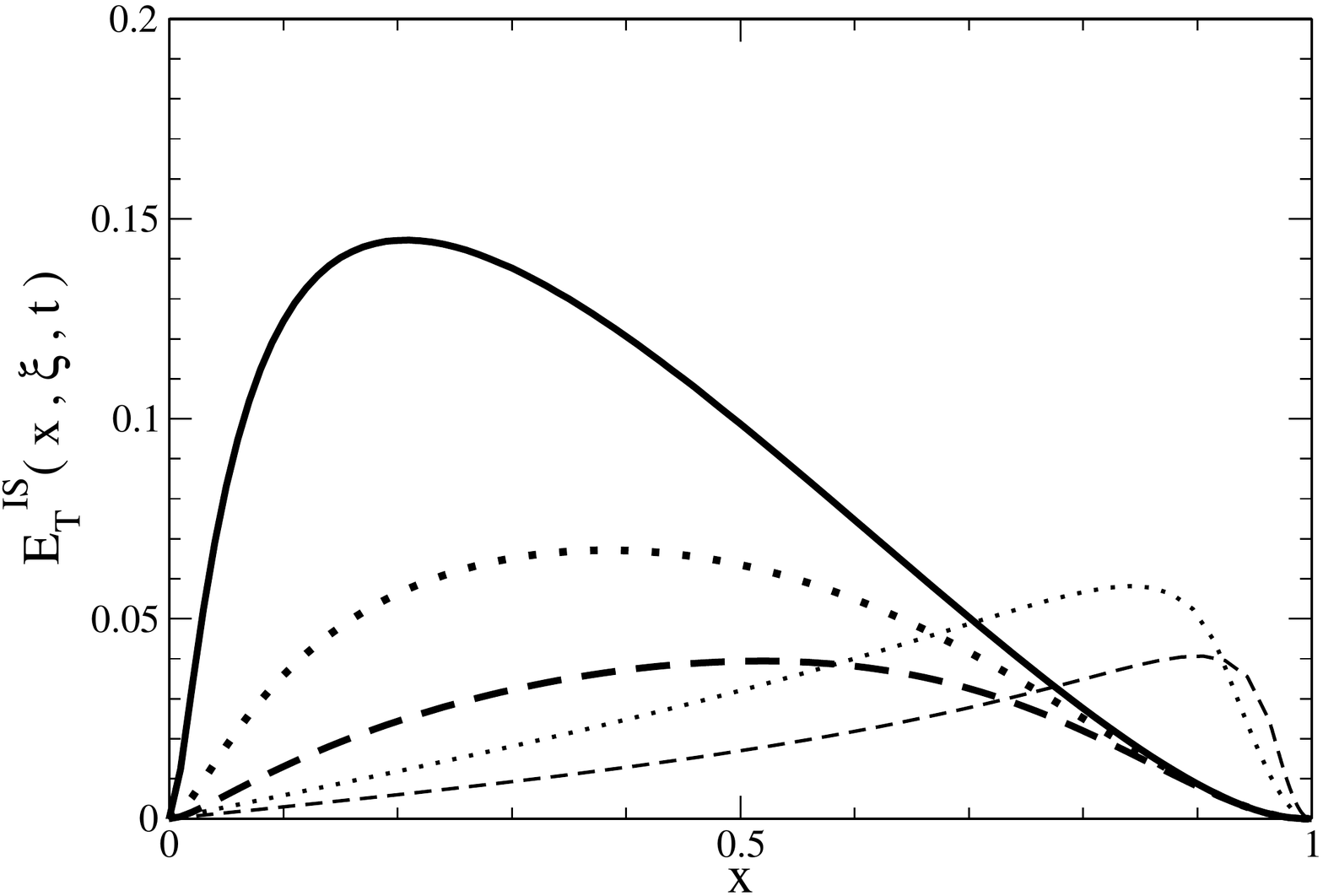} \hspace{-0.1cm}
\includegraphics[width=6.0cm]{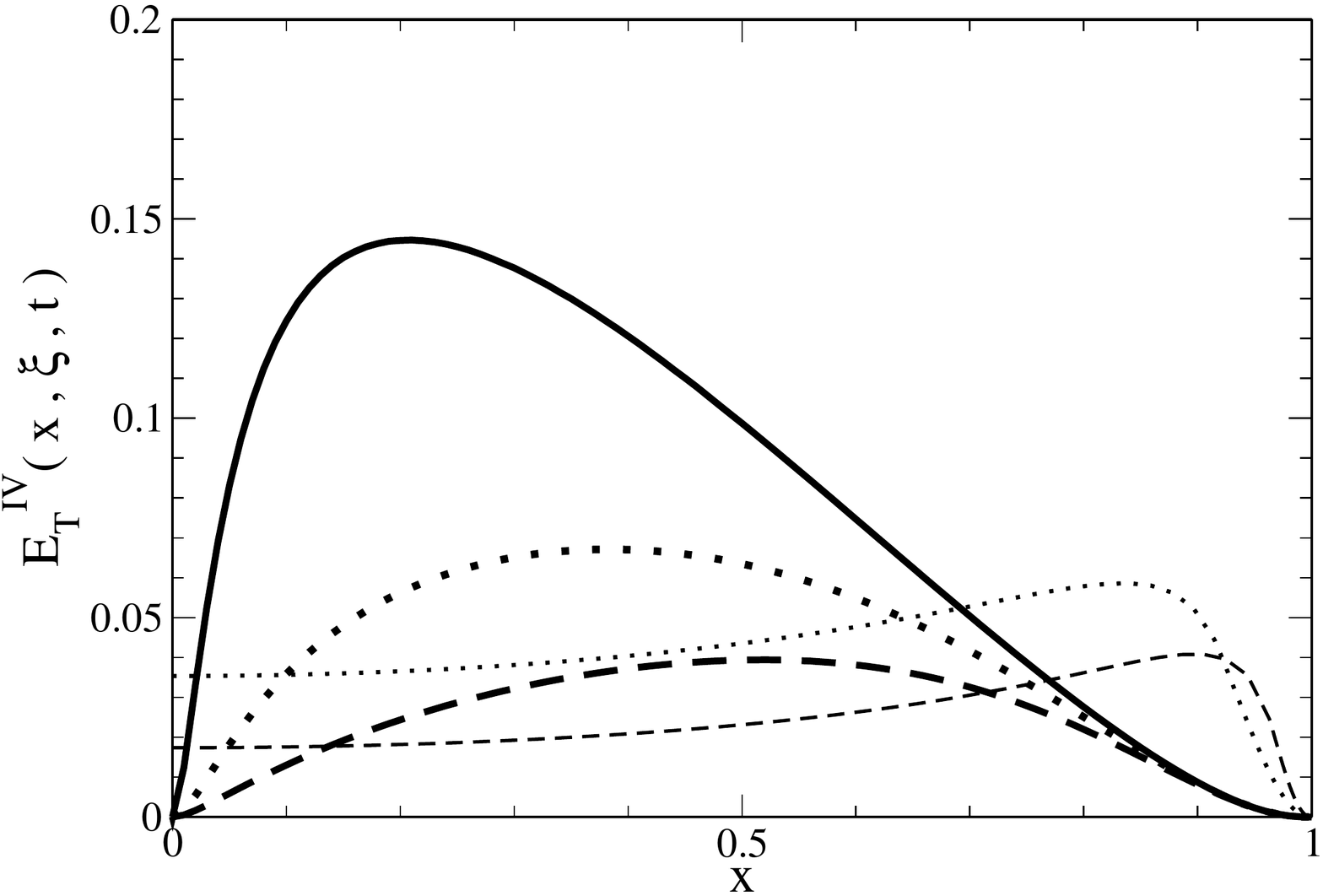}
\end{center}
\caption{Preliminary isoscalar and isovector tensor GPD's for a charged pion, within CCQM, for positive
$x$. The behavior for negative values of $x$ can be obtained by recalling Eq.
(\ref{sym}). 
Thick solid line: $\xi=0~$ and $t=0$.
Thick dotted line: $\xi=0~$ and $t=-0.4\,GeV^2$.
Thick dashed line: $\xi=0~$ and $t=-1\,GeV^2$. Thin dotted line: 
$\xi=0.96$ and $t=-0.4\,GeV^2$. Thin dashed line:  
$\xi=0.96$ and $t=-1\,GeV^2$.}
\end{figure}

From the tensor GPD, one can calculate the corresponding GFF, see 
Eq. (\ref{gff}). We have  started (see Ref. \cite{Pace12} for a preliminary
presentation) the investigation by evaluating 
$B_{1,0}(t)$ and  
$B_{2,0}(t)$, and  comparing our
results with the  
 lattice  data \cite{Brom07,Brom08,Bromth},   as well as with other models 
 like the chiral quark model of
Ref. \cite{Broni10} and the instanton model of Ref. \cite{Nam11}. The complete analysis
will be presented elsewhere \cite{PRS12}. For a better comparison between our calculations of GFF's and the
lattice results, 
we have  also evaluated $B_{1,0}(t)$ and $B_{2,0}(t)$ both at the physical value
of the pion mass, $m_\pi=140$ MeV, and at the actual 
lattice value, $m^{lat}_\pi=600$ MeV,  tuning the CQ mass according to  
the  Chiral Perturbation Theory (see, e.g. Ref. \cite{Cloet}), namely
\be
 m_q=m^{Phys}_q + \left [ {m^2_\pi \over (m^{Phys}_\pi)^2} -1\right] m_c^{Phys}
\label{runm}\ee
 where $m^{Phys}_q$ and $m^{Phys}_c$ are the constituent quark mass and the current quark mass 
at the physical pion mass, respectively. 
In our approach the values  $m^{Phys}_q=220$ MeV and $m_c^{Phys}=4\pm 1$ MeV 
 have been taken. Notice that  lattice data were extrapolated to the physical
 pion mass by using a
proper chiral formula \cite{Brom08,Bromth}. In particular, we can anticipate that
the behavior of 
 $B_{1,0}(t)/B_{1,0}(0)$ fully agrees with both i) the lattice data \cite{Brom08}
  and ii) the calculations from $\chi$QM
 \cite{Broni10} and instanton model \cite{Nam11}, while for 
 $B_{2,0}(t)/B_{2,0}(0)$ the comparison improves as the physical pion mass is
 approached. It is worth noting that the GFF's has been rescaled by their own
 value at $t=0$, in order to get rid off the effect of the QCD evolution on the
 tensor GPD
 (cf \cite{Broni10} and \cite{Nam11}). Furthermore, since the tensor charges of the
 pion are
 attracting an increasing interest (see also Ref. \cite{marti}, where  
 $B_{1,0}(0)$ is taken as  a check for 
 the lattice evaluation of other observables),
  it is important to carefully investigate 
  $B_{n,0}(0)$, for $n=1,2$, as a function of a running $m_\pi$, for getting 
  information on
  their  chiral limit (i.e. the relevant limit for lattice calculations). 
  In Fig. 3, the preliminary results of CCQM for both   
  $B_{1,0}(0,m_\pi)$ and  $B_{2,0}(0,m_\pi)$ are shown;  it is understood that in
  our analytic model,  $B_{n,0}(0, m_\pi)$ is evaluated by changing 
  the values of $m_\pi$ and  $m_q$, according to Eq. (\ref{runm}).
  It is worth noting that the linear behavior as a function of $m_\pi$,
  starts  for $B_{2,0}(0,m_\pi)$ earlier than for  $B_{1,0}(0,m_\pi)$, and  in
  particular,
   for $m_\pi\le 300$ MeV one has
   $B_{1,0}(0,m_\pi)\sim 1.13~m_\pi$ and $B_{2,0}(0,m_\pi)\sim
   0.74~m_\pi$.

To accomplish a meaningful comparison with the lattice data, one has to evolve
the CCQM results up to the scale $\mu=2$ GeV, the scale of the lattice data.
 According to \cite{Broni10}, the evolution should affect
$B_{2,0}(0)$ more than $B_{1,0}(0)$ (almost a factor of two for $\mu_0=320$ MeV).   
\begin{figure}

\vspace{-3cm}
\centering
\includegraphics[width=6cm]{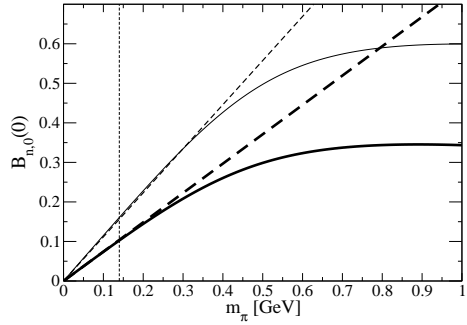} 
\caption{ Preliminary $B_{n,0}(0)$ vs $m_\pi$, for $n=1,2$. 
Thin solid line: CCQM $B_{1,0}(0,m_\pi)$. Thin dashed
line: $B_{1,0}^\chi(0,m_\pi)=1.13~m_\pi$.
Thick solid line: CCQM  $B_{2,0}(0,m_\pi)$. Thick dashed line: 
$B_{2,0}^\chi(0,m_\pi)=0.74~m_\pi$. As a  guide, the physical pion mass is
indicated by a vertical dashed line.}
\label{mass0}
\end{figure}
  Though
the complete investigation will be
reported elsewhere \cite{PRS12}, we can anticipate here that
for   $n=1$, the CCQM tensor charge    is lower than the lattice  one presented in
in Ref. \cite{Brom08}, that in turn is bigger than the recent lattice data of Ref.
\cite{marti}. Furthermore,
 the comparison with the $\chi$QM calculation \cite{Broni10},
  in the limit of 
$m_\pi\to 0$, seems reasonable, since $B^{\chi QM}_{1,0}(0)\sim 1.33~m_\pi$ and 
$B^{\chi QM}_{2,0}(0)\sim 0.47~m_\pi$ at the quark-model scale $\mu_0=320$ MeV.

The  calculation of the GFF's, allows one to investigate the probability 
density $\rho_n({\bf b}_\perp,{\bf s}_\perp)$ for a transversely-polarized
u-quark
 (cf Eq. (\ref{rhodef})) so that one can address the IPS structure of the pion.
  Table I shows  preliminary
results (still without evolution effects) for  the
average transverse shifts when the   quark is polarized along the $x$-axis, 
i.e. ${\bf s}_\perp\equiv \{1,0\}$. The shift for a given $n$ is
 \be
 \langle b_y\rangle_n= {\int d{\bf b}_\perp~b_y~ \rho_n({\bf b}_\perp,{\bf s}_\perp)
 \over \int d{\bf b}_\perp~ \rho_n({\bf b}_\perp,{\bf s}_\perp)}= {1 \over 2 m_\pi}~
 {B_{n,0}(t=0)\over 
 A_{n,0}(t=0)}
 \ee
  \begin{table}
\centering
\caption{Mean shifts along the direction perpendicular to the u-quark 
transverse  polarization, ${\bf s}_\perp \equiv \{1,0\}$, for $n=1,2$.}
\begin{tabular} { |c | c  | c |}
\hline
 ~        & CCQM - $m_\pi = 140$ MeV & Lattice \cite{Brom08}\\
\hline \hline
$\langle b_y\rangle_1$  & 0.113 ~fm& 0.151 $\pm $ 0.024  ~fm\\
\hline
$\langle b_y\rangle_2$  & 0.092 ~fm  & 0.106 $\pm$ 0.028  ~fm\\
\hline
\end{tabular}
\end{table} 
 The values shown in Table I clearly demonstrate the dipole-like distortion of 
the transverse density
in a direction perpendicular to the quark polarization, pointing to a non
trivial correlation between the orbital angular momenta  and the  spin
of the constituents inside a pseudoscalar hadron
 (see, e.g., Ref. \cite{Bur08} and references quoted  therein).
\section{Conclusions and Perspectives}
 A simple, fully  covariant constituent quark model  
 has been exploited  for investigating the phenomenology of the leading-order 
 Generalized Parton
 Distributions of the pion. The model has been already applied to the
 vector GPD \cite{Frede09,Frede11}, and in the present work some preliminary results 
 for the tensor GPD
 have been illustrated.
 The main ingredients of the approach are
 i) an Ansatz for the  Bethe-Salpeter amplitude  and ii) the
 generalization of the Mandelstam formula, applied in the seminal work of 
 Ref.  \cite{Mandel} to matrix elements
 of the em current operator between states of  a relativistic composite system.
   
The CCQM tensor GPD, $E_{\pi T}(x,\xi,t)$, favorably compares with the one obtained
within the 
 Light-front Hamiltonian dynamics (see \cite{Frede11,Pace12}) and this represents an important
 test of the CCQM, as well as the analysis of the tensor charges by varying the
 pion mass (see Fig. 3), relevant for the chiral extrapolation of the lattice 
 results.  Finally the first evaluation of the
 transverse shifts of the probability distribution of a  
 transversely-polarized quark has been given in Table I, indicating an encouraging comparison
 with the lattice data.
 
 In order to construct a more and more reliable phenomenological tool, the  CCQM
  study of the pion GPD will be improved by including 
  the 
  evolution with the energy scale, and by  
    developing  new  Ansatzes of the Bethe-Salpeter amplitude, that allow one to
    increase the 
  dynamical content, for instance within the Nakanishi perturbation integral
  representation of the quark-hadron vertex (see, e.g., Ref. \cite{Frede12}).

\end{document}